\documentclass[conference]{IEEEtran}
\usepackage{tabularx,makecell,booktabs,array,graphicx}
\IEEEoverridecommandlockouts

\usepackage[final]{microtype}
\usepackage[utf8]{inputenc}
\usepackage{graphicx}
\usepackage{amsmath,amssymb}
\usepackage{upgreek}
\usepackage{siunitx} 
\usepackage{algorithm}
\usepackage{algpseudocode}
\usepackage{tabularx}
\usepackage{booktabs}
\usepackage{makecell}
\usepackage{cite}
\usepackage{amsmath,amssymb,amsfonts}
\usepackage{graphicx}
\usepackage{textcomp}
\usepackage{xcolor}
\usepackage{tabularx,makecell,booktabs,array,graphicx}
\PassOptionsToPackage{hyphens}{url} 
\usepackage{subcaption}
\usepackage{enumitem}
\usepackage{hyperref}
\setlist[itemize]{leftmargin=*}
\def\BibTeX{{\rm B\kern-.05em{\sc i\kern-.025em b}\kern-.08em
    T\kern-.1667em\lower.7ex\hbox{E}\kern-.125emX}}
\begin{document}

\makeatletter
\def\bstctlcite#1{\@bsphack
  \@for\@citeb:=#1\do{\edef\@citeb{\expandafter\@firstofone\@citeb}%
    \if@filesw\immediate\write\@auxout{\string\citation{\@citeb}}\fi}%
  \@esphack}
\makeatother
\bstctlcite{IEEEexample:BSTcontrol}

\title{
Six Times to Spare: Characterizing GPU-Accelerated 5G LDPC Decoding for Edge-RSU Communications
\thanks{This work was supported, in part, at Clemson University by the State of South Carolina through funding for the Battelle Savannah River Alliance Workforce Development Program. It is also funded by the National Science Foundation under Grant Numbers CNS-2202972, CNS- 2318726, and CNS-2232048.}}

\author{
  \IEEEauthorblockN{
    Ryan Barker\IEEEauthorrefmark{1},
    Julia Boone\IEEEauthorrefmark{1},
    Tolunay Seyfi\IEEEauthorrefmark{1},
    Alireza Ebrahimi Dorcheh\IEEEauthorrefmark{1},
    Fatemeh Afghah\IEEEauthorrefmark{1},
    Joseph Boccuzzi\IEEEauthorrefmark{4}
  }
  \IEEEauthorblockA{
    \IEEEauthorrefmark{1}Holcombe Department of Electrical and Computer Engineering, Clemson University, Clemson, SC, USA\\
    Emails: \{rcbarke, jcboone, tseyfi, alireze, fafghah\}@clemson.edu
  }
  \IEEEauthorblockA{
    \IEEEauthorrefmark{4}NVIDIA Corporation, Santa Clara, CA, USA\\
    Email: jboccuzzi@nvidia.com
  }
}

\maketitle

\begin{abstract} 

Ultra-reliable low-latency vehicular communications (URLLC) require sufficient physical-layer (PHY) compute headroom at the network edge, where roadside units (RSUs) and compact next-generation base stations (gNBs) must meet strict timing constraints while co-hosting higher-layer services. In 5G New Radio (5G NR), low-density parity-check code (LDPC) decoding is a latency-sensitive iterative PHY workload whose cost scales with both workload parallelism and decoder iteration budget, making it a potential bottleneck on general-purpose central processing units (CPUs). This paper presents a reproducible, telemetry-backed microbenchmark derived from the Sionna LDPC5G baseline to characterize the compute headroom obtained through graphics processing unit (GPU) offload on compact heterogeneous edge platforms. We evaluate decoder behavior across multiple processor architectures and a wide range of batch sizes and iteration counts, with emphasis on dense operating regimes relevant to edge provisioning. Results show that GPU acceleration substantially increases LDPC throughput, reduces amortized decode service time, and shifts compute pressure away from the CPU, thereby improving the feasibility of meeting edge-RSU timing budgets under heavy parallel workloads. These findings indicate that GPU offload can provide substantial spare PHY compute margin for compact vehicular edge platforms, making dense decode workloads more practical within realistic edge power and timing constraints.

\end{abstract}


\section{Introduction: Edge-RSU Compute Headroom for URLLC Vehicular Communications}
\label{sec:intro}

Autonomous vehicle (AV) and Vehicle-to-Everything (V2X) communication stacks are safety-critical systems that must remain dependable under both nominal operation and rare stress conditions such as degraded infrastructure or bursty demand. Recent robotaxi service interruptions highlight that large-scale failures can arise from systemic and operational factors even when the physical layer itself remains functional. These realities motivate robust and scalable \emph{edge autonomous infrastructure} for vehicular deployments in next-generation wireless networks~\cite{roy2025_waymo_outage}.

A key implication is that roadside units (RSUs) and compact edge base stations must deliver Ultra-reliable low-latency  communications (URLLC)-grade responsiveness while simultaneously hosting compute-intensive higher-layer functions such as coordination, cooperative perception, and policy control. This pushes vehicular edge architectures toward compact heterogeneous compute nodes (e.g., NVIDIA DGX Spark) and careful real-time resource provisioning~\cite{liu2021vec_survey,butcher_boccuzzi_2019_dell_ran,10978559}. Within the radio stack, forward-error-correction (FEC) decoding remains a dominant compute kernel; in 5G New Radio (5G NR) pipelines, low-density parity-check code (LDPC) decoding is both latency-sensitive and throughput-intensive, making it a natural candidate for accelerator offload in collocated deployments.

Despite extensive work on accelerated LDPC implementations, publicly reproducible \emph{system-oriented} studies of new-radio-like LDPC decoding on compact edge nodes remain limited, particularly under heavy parallel workloads intended to probe saturation and compute headroom. To address this gap, we develop a reproducible microbenchmark harness based on the Sionna LDPC baseline~\cite{sionna_ldpc_api}, enabling controlled central processing unit to graphics processing unit (CPU--GPU) comparisons with consistent inputs and telemetry-backed reporting across two hardware classes: a compact DGX Spark edge node and a commercial-off-the-shelf (COTS) workstation reference system.\footnote{\url{https://github.com/rcbarke/six-times-to-spare}} Our focus is on measurement methodology and translating decoder timing into practical provisioning insights for edge RSU/next-generation NodeB (gNB) deployments.

To support this analysis, we characterize three operating regions: launch-limited small batches, a GPU ramp-up regime where parallel efficiency improves with increasing the number of codewords decoded per batch, and a dense steady-state regime in which accelerator service time stabilizes. This structure preserves edge-relevant headroom as the central key performance indicator (KPI) while also revealing baseline-to-saturation behavior.

\begin{itemize}
\item \textbf{Reproducible LDPC decode benchmark:} We construct an NR-like LDPC decoding benchmark from the Sionna LDPC baseline~\cite{sionna_ldpc_api}, executed with identical inputs and execution graphs across CPU and GPU back ends, isolating compute.

\item \textbf{Two-regime codeword ablation:} We sweep codeword-level batch parallelism over a baseline regime of 1 to 1024 codewords, in powers of two, and a dense regime of 2048 to 20480 codewords, in increments of 2048 codewords. For every batch size, we sweep belief-propagation iteration budget from 4 to 22 iterations to characterize the throughput--latency tradeoff under realistic decoder workload ramp-up. We report batch latency, throughput, and amortized service time as a ratio of batch latency to batched codewords.

\item \textbf{Telemetry-backed edge characterization:} System telemetry (CPU utilization, GPU utilization, and power) quantifies compute offload, resource rebalancing, and decoder headroom on DGX Spark under identical workloads.

\item \textbf{Edge versus workstation comparison:} Results are contrasted with a COTS i9-14900K/RTX~4090 system to contextualize absolute acceleration while preserving focus on compact edge platforms for RSU/gNB deployment.
\end{itemize}

\section{Background and Related Work}
\label{sec:related-work}

\subsection{URLLC and vehicular edge context}
\label{subsec:urllc-v2x}

Ultra-reliable low-latency communication (URLLC) targets stringent latency–reliability envelopes for safety-critical traffic, motivating cross-layer mechanisms such as short transmission intervals, diversity, and careful processing budgets~\cite{popovski2018urllc_building_blocks,qualcomm2017_urllc_systems}. In vehicular networks, these requirements arise in V2X use cases ranging from cooperative awareness to coordinated maneuvers, where tail latency guarantees and predictable responsiveness can be as important as average throughput. 5G NR-V2X extends the NR framework toward such scenarios through both network-assisted and direct communication modes, requiring coordinated radio configuration, precise scheduling, and system implementation choices~\cite{bagheri2021nr_v2x}.

A key implication is that URLLC performance cannot be treated purely as an air-interface problem. Roadside units (RSUs) and edge base stations increasingly host application logic and network services close to the roadway to reduce round-trip latency and improve robustness under load. Architectures such as mobile edge computing (MEC) and software-defined vehicular networking (SDVN) explicitly pursue this approach by colocating compute and control with the access network~\cite{huang2017mec_5g_sdvn,malinverno2020mec_collision_avoidance,barmpounakis2020vru_safe}. As a result, vehicular edge platforms must be provisioned as heterogeneous shared resources where compute headroom and contention become central design concerns alongside radio performance~\cite{qualcomm2017_urllc_systems,liu2021vec_survey}.

\subsection{Edge workloads beyond communications}
\label{subsec:edge-comm}

Beyond communications processing, RSU and edge deployments increasingly support higher-layer intelligence that benefits from proximity and multi-vehicle aggregation. Cooperative perception is a representative example in which edge-assisted systems fuse sensor observations from multiple vehicles to extend sensing range and mitigate occlusions~\cite{zhang2021emp}. These workloads involve latency-sensitive tasks such as inference, fusion, and filtering and therefore compete with baseband processing for shared compute resources, reinforcing the need to quantify how much physical-layer processing budget can be reclaimed through acceleration.

\subsection{Network reliability}
\label{subsec:network-reliability}

In URLLC-oriented 5G/6G systems, reliability is achieved through a combination of physical-layer robustness and fast link-layer control loops operating under strict timing constraints. Hybrid automatic repeat request (HARQ) couples forward-error correction with retransmissions: a transport block is transmitted using a selected modulation and coding scheme, decoded at the receiver, and acknowledged via ACK/NACK feedback within a bounded processing timeline. If decoding fails, retransmissions with incremental redundancy allow the receiver to combine log-likelihood ratios (LLRs) across HARQ rounds to improve decoding probability~\cite{popovski2018urllc_building_blocks,qualcomm2017_urllc_systems}. 

For an edge RSU or compact gNB, these mechanisms translate directly into compute deadlines. The receiver must complete demodulation, LLR generation, and Forward Error Correction (FEC) decoding rapidly enough to support HARQ feedback and scheduling decisions within the transmission interval. Vehicular deployments can further introduce bursts of simultaneous demand from multiple concurrent user equipment (UE) connections and HARQ processes, creating intervals where many code blocks must be decoded within a single slot budget~\cite{bagheri2021nr_v2x}. This motivates treating decoder execution as a batched, highly parallel service when evaluating compute headroom on an edge node.

\subsection{LDPC decoding and acceleration}
\label{subsec:ldpc-hpc}

Within this reliability loop, LDPC decoding is a critical compute kernel because it is invoked per code block and per HARQ round, directly influencing whether timing deadlines can be met under load. LDPC codes are capacity-approaching channel codes decoded via iterative message passing~\cite{gallager1962ldpc}. In 5G NR they are standardized for data channels, with decoding cost scaling with both workload parallelism and the belief-propagation iteration budget. To this end, decoders employ specification-defined code construction and rate-matching procedures that interact with retransmission behavior and soft combining across HARQ rounds~\cite{3gpp38212}. 

Prior work has demonstrated that GPU-based LDPC implementations can achieve multi-gigabit throughput and has explored design tradeoffs for modern parallel processors~\cite{dai2022multigbps,lu2023gpuassisted}. Heterogeneous computing has likewise been proposed for RAN workloads in compact edge platforms~\cite{butcher_boccuzzi_2019_dell_ran}. However, publicly reproducible system-oriented studies that connect LDPC decoder behavior to edge-RSU provisioning questions—such as HARQ timing margins, concurrency headroom, and resource contention with colocated edge workloads—remain limited, motivating the measurement-driven approach of this work.

\section{System Model and Digital Signal Processing}
\label{sec:system-model}

\begin{figure}[t]
  \centering
  \includegraphics[width=\linewidth]{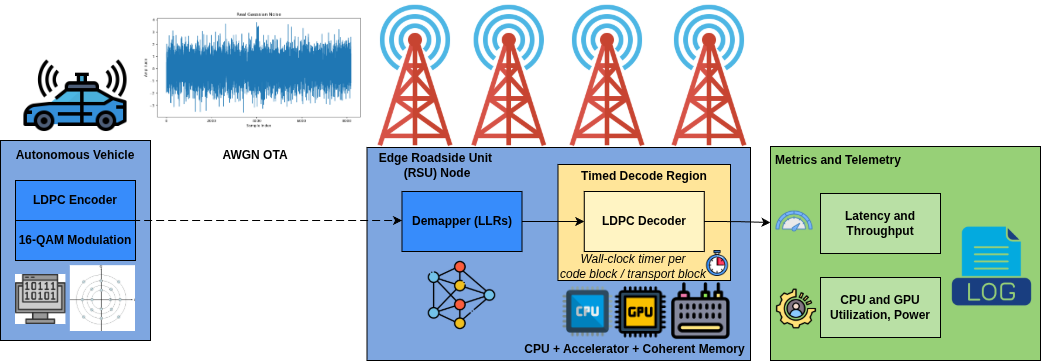}
  \caption{{\small Systems model and benchmark PHY workflow for edge-RSU LDPC decode characterization. An autonomous vehicle generates coded 16-QAM symbols and an additive Gaussian white noise channel (AWGN) produces received samples that are demapped into  LLRs at an edge RSU node. The RSU is modeled as a compact heterogeneous platform (CPU + accelerator with coherent memory). The LDPC decoder is the \emph{timed} region of the pipeline, while we log latency/throughput metrics and system telemetry (CPU/GPU utilization and power) for each configuration.}}
  \label{fig:method-system-model}
\end{figure}

Figure~\ref{fig:method-system-model} models an \emph{edge base-station node} co-located with a RSU or small-cell gNB, where baseband processing must meet URLLC-grade responsiveness while sharing compute resources with higher-layer edge applications. The node comprises (i) a general-purpose CPU for control and orchestration, (ii) an on-node accelerator (e.g., a GPU-class single-instruction multiple-data / single-instruction multiple-thread (SIMD/SIMT) device) for parallelizable kernels, and (iii) shared memory accessible to both compute domains. 


Within this node, the FEC stage maps soft information to decoded bits. Here, \(E(\cdot)\) denotes the LDPC encoder, \(M(\cdot)\) the 16-QAM mapper, \(\mathcal{L}(\cdot)\) the soft demapper, and \(D(\cdot; I)\) the iterative LDPC decoder with iteration budget \(I\); \(k\) and \(n\) denote the information and coded block lengths, respectively, and \(N_{\mathrm{cw}}\) is the number of codewords decoded in parallel. For an LDPC code of length \(n\) and dimension \(k\) (rate \(R=k/n\)), the decoder computes \(\hat{b} = D(L; I) \in \{0,1\}^{k}\) from an LLR vector \(L \in \mathbb{R}^{n}\) using \(I\) belief-propagation iterations. To represent RSU operating conditions with multiple simultaneous code blocks (e.g., across users or HARQ processes), we model the workload as batched decoding over \(N_{\mathrm{cw}}\) codewords with input tensor \(L \in \mathbb{R}^{N_{\mathrm{cw}}\times n}\) and outputs \(\hat{B} \in \{0,1\}^{N_{\mathrm{cw}}\times k}\).

We instantiate this model on a compact heterogeneous edge platform (DGX Spark), NVIDIA's representative of emerging CPU--accelerator nodes designed for the edge, from their larger Deep GPU Xcceleration (DGX) high-performance computing (HPC) line. DGX Spark integrates a 20-core Arm Grace CPU and an on-package 1\,PFLOP Blackwell-class GPU within a coherent System-on-Chip (SoC), sharing 128\,GB of LPDDR5x memory via NVLink-chip-to-chip (C2C). Compared with conventional x86 hosts, Grace’s Arm-based design emphasizes performance-per-watt, multicore scalability, and coherent CPU–GPU memory access rather than simply maximizing host-side clock frequency. This enables high-bandwidth exchange between CPU control logic and accelerator-executed compute graphs and reduces tensor overhead of moving tensors such as $\mathbf{L}$ compared to peripheral component interconnect express (PCIe)-attached discrete units.

We also evaluate a conventional COTS workstation configuration for comparison. DGX Spark uses a Grace CPU and GB10 GPU within a coherent shared-memory SoC, whereas the COTS system uses an Intel i9-14900K CPU and an NVIDIA RTX~4090 GPU connected through PCIe with separate random access memory (RAM) and video random access memory (VRAM). DGX Spark represents the deployment target for compact power-limited edge regimes, while the COTS system serves only as a workstation-class upper-bound reference.

We benchmark an NR-like link-level PHY chain workload implemented using Sionna's 5G-compliant LDPC components~\cite{sionna_ldpc_api,hoydis2023sionna,horrocks2022jumpstarting} consistent with 3GPP NR channel coding procedures~\cite{3gpp38212}. Information bits $\mathbf{b}\in\{0,1\}^{k}$ are encoded to a codeword $\mathbf{c}=\mathcal{E}(\mathbf{b})\in\{0,1\}^{n}$ with $(k,n)=(512,1024)$ ($R=1/2$) and mapped to 16-QAM symbols $\mathbf{x}=\mathcal{M}(\mathbf{c})$. 
AWGN channel transmission produces \(y = x + w\) with \(w \sim \mathcal{CN}(0, N_0 \mathbf{I})\), and a soft demapper generates LLRs \(L = \mathcal{L}(y, N_0) \in \mathbb{R}^{n}\) used as decoder inputs.

In this notation, \(N_0\) is the AWGN noise parameter supplied to the demapper, and the benchmark holds \((k,n)\), modulation order, and channel model fixed so only decoder batch parallelism and iteration count vary across experiments. 
Our controlled chain provides repeatable LLR tensors while holding the coded-modulation structure, determining LDPC compute. Thus, \(N_{\mathrm{cw}}\) should be interpreted not as a protocol field, but as an abstracted concurrent decode demand arising from multiple users, code blocks, or HARQ processes at the edge.

\section{Telemetry-Backed LDPC Decode Microbenchmark Protocol}
\label{sec:methodology}

Building on the edge-node model and NR-like DSP chain in Section~\ref{sec:system-model}, we design a microbenchmark that isolates \emph{LDPC decoding} as the timed kernel while keeping all other processing fixed or outside the timed region. For a batched LLR tensor \(L \in \mathbb{R}^{N_{\mathrm{cw}}\times n}\), the decoder computes \(\hat{B} = D(L; I) \in \{0,1\}^{N_{\mathrm{cw}}\times k}\) using \(I\) belief-propagation iterations. To ensure CPU--GPU comparability, the same TensorFlow compute graph derived from the Sionna LDPC5G baseline is executed on either device via explicit placement, treating the hardware binding as the only experimental variable.

The PHY workload remains fixed across experiments. Information bits are encoded using the rate-$1/2$ NR-like LDPC chain $(k,n)=(512,1024)$, mapped to 16-QAM symbols, transmitted through AWGN, and soft-demapped to produce decoder-input LLRs. Two stress parameters are then varied: batch parallelism $N_{\mathrm{cw}}$ and iteration count $I$. We sweep $N_{\mathrm{cw}}\in\{1,2,4,\ldots,1024\}$ in a \emph{baseline regime} and $N_{\mathrm{cw}}\in\{2048,4096,\ldots,20480\}$ in a \emph{dense regime}, with $I\in\{4,6,\ldots,22\}$. All other chain parameters remain fixed, so \(N_{\mathrm{cw}}\) controls codeword-level parallel workload size while \(I\) controls per-codeword decoder effort. The baseline regime exposes launch and ramp-up behavior, whereas the dense regime characterizes the steady-state operating region relevant for edge provisioning.

For each configuration $(N_{\mathrm{cw}},I)$, a single LLR tensor $\mathbf{L}$ is generated and reused across all timed decodes to eliminate stochastic channel variation. The tensor is materialized on the target device and a warm-up decode is executed to amortize graph tracing and allocator setup. We then perform $M=10$ timed decodes and summarize them as one \emph{outer trial}. The entire procedure is repeated for $R=10$ trials per configuration to reduce the impact of OS jitter and background activity.

The primary latency metric is the batch decode latency
\[
t_{\mathrm{dec}} \triangleq \frac{1}{M}\sum_{m=1}^{M} t^{(m)}_{\mathrm{dec}},
\]
measured with a synchronized high-resolution timer. From this we derive information throughput
\[
T_{\mathrm{thr}} = \frac{N_{\mathrm{cw}}k}{t_{\mathrm{dec}}}
\]
(bits/s) and amortized per-codeword service time
\[
t_{\mathrm{cb}} = \frac{t_{\mathrm{dec}}}{N_{\mathrm{cw}}},
\]
interpreted as an effective batch-parallel service time rather than a single-codeword latency.

System telemetry is collected in parallel with timing. GPU utilization and power are sampled periodically (approximately 1\,Hz), and process-level CPU usage is converted to \emph{active-core equivalents} via $C_{\mathrm{cpu}}\approx U_{\mathrm{cpu}}/100$. Because the implementation uses high-level Sionna/TensorFlow operators rather than hand-tuned CUDA kernels, the measured speedups represent conservative lower bounds on accelerator performance.

Reported results aggregate along different axes. Figure~\ref{fig:throughput_scaling_codewords} shows throughput versus $N_{\mathrm{cw}}$ aggregated across iteration counts, while Figure~\ref{fig:throughput_scaling_dense} shows throughput versus iteration count aggregated across dense-regime batch sizes. Table~\ref{tab:cots_spike_ablation} reports a focused COTS spike ablation at pre-selected \((N_{\mathrm{cw}}, I)\) operating points, with throughput shown as observed range across repeated runs. Table~\ref{tab:per_cb_latency} reports dense-regime service time $t_{\mathrm{cb}}$ averaged over the ten batch sizes and $R=10$ trials.

The objective is not to emulate a specific scheduler instance but to \emph{saturate the compute substrate} and characterize decoder headroom. Large $N_{\mathrm{cw}}$ exposes steady-state accelerator behavior, while the baseline regime reveals launch-limited and ramp-up regions. Sweeping $I$ captures the primary algorithmic knob controlling decode cost. If the platform maintains URLLC-relevant timing margins under this intentionally stressed regime, then typical operating points with smaller batches or fewer iterations should admit additional headroom for the remainder of the PHY and concurrent edge workloads.

\section{Results and Evaluation}
\label{sec:results}

This section summarizes the throughput, service-time, and resource-usage outcomes of the telemetry-backed LDPC5G decode study described in Sections~\ref{sec:system-model}--\ref{sec:methodology}. We report information-bit throughput as $T_{\mathrm{thr}} \triangleq \frac{N_{\mathrm{cw}} k}{t_{\mathrm{dec}}}$, where $t_{\mathrm{dec}}$ is the measured batch decode latency for a tensor containing $N_{\mathrm{cw}}$ codewords of dimension $k$. Because our study includes two CPU/GPU pairs, we report \emph{paired} throughput speedups within each platform: $S_{\mathrm{thr}}^{(\mathrm{DGX})} \triangleq \frac{T_{\mathrm{thr}}^{(\mathrm{GB10})}}{T_{\mathrm{thr}}^{(\mathrm{Grace})}}$ and $S_{\mathrm{thr}}^{(\mathrm{COTS})} \triangleq \frac{T_{\mathrm{thr}}^{(\mathrm{4090})}}{T_{\mathrm{thr}}^{(\mathrm{i9})}}$. This paired notation is important because the DGX Spark and COTS platforms serve different roles in the paper: DGX Spark is the edge target, while COTS provides a workstation-class comparison for absolute performance.

\begin{figure*}[t]
  \centering

  \begin{subfigure}[t]{0.4\textwidth}
    \centering
    \includegraphics[width=\linewidth]{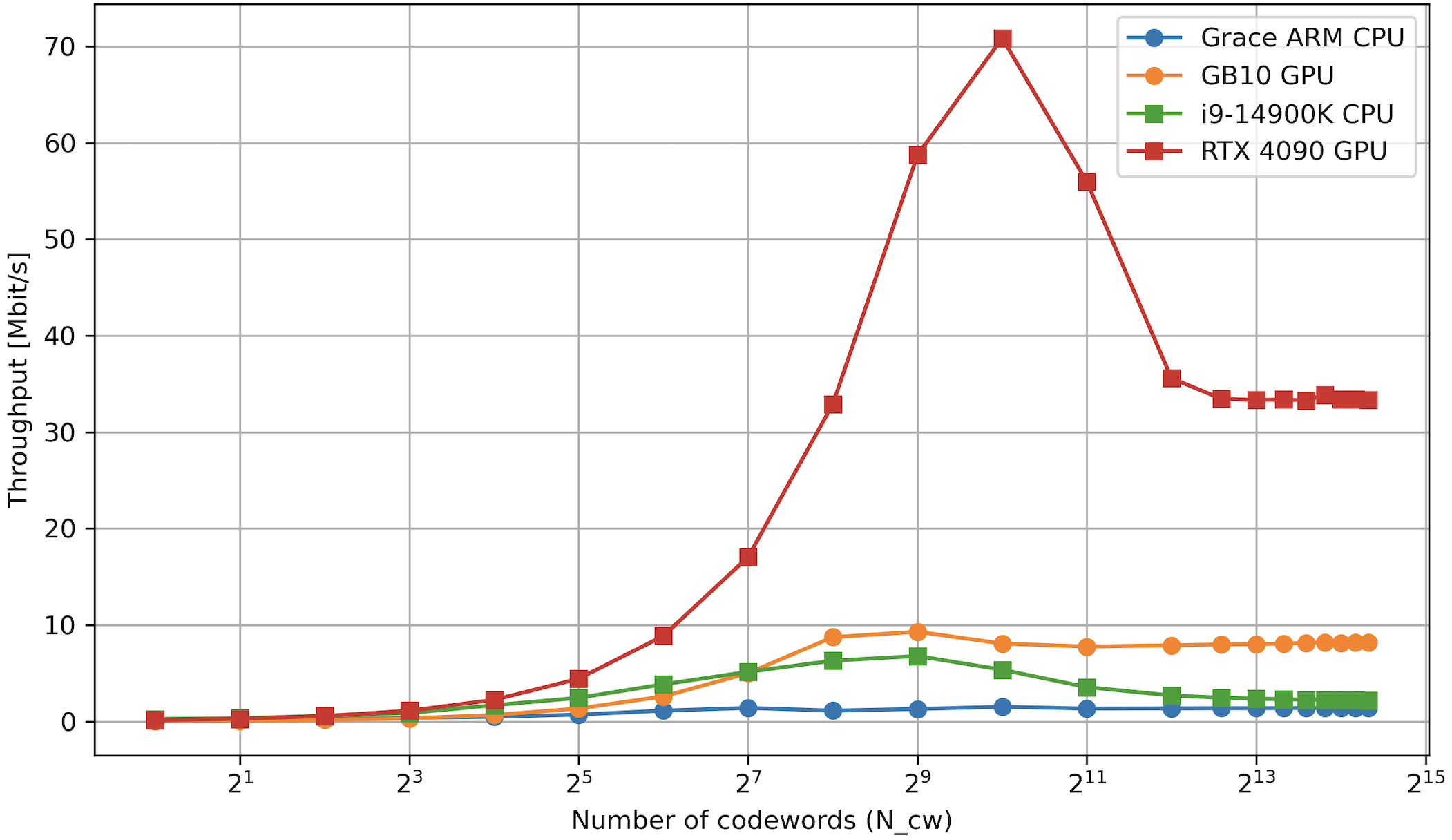}
    \caption{Throughput versus number of codewords $N_{\mathrm{cw}}$, aggregated over $I\in\{4,6,\ldots,22\}$.}
    \label{fig:throughput_scaling_codewords}
  \end{subfigure}
  \hfill
  \begin{subfigure}[t]{0.4\textwidth}
    \centering
    \includegraphics[width=\linewidth]{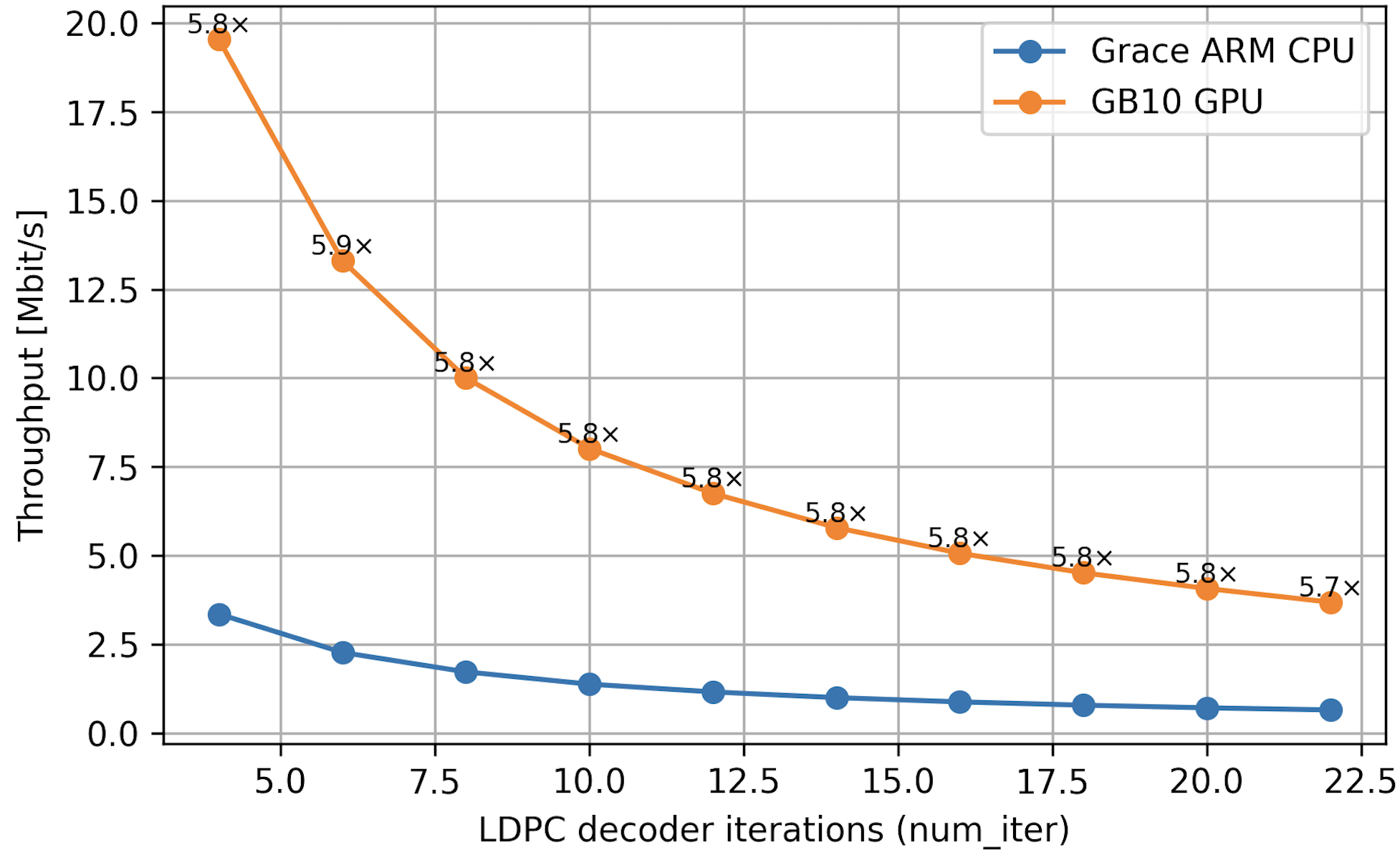}
    \caption{DGX Spark throughput versus decoder iterations in the dense regime.}
    \label{fig:throughput_scaling_dense}
  \end{subfigure}

  \caption{{\small LDPC5G throughput scaling across hardware and operating regime. \textbf{Left:} throughput versus number of codewords $N_{\mathrm{cw}}$, aggregated over $I\in\{4,6,\ldots,22\}$, for all four devices; the $N_{\mathrm{cw}}$ axis is shown on a $\log_2$ scale so that both the baseline regime ($N_{\mathrm{cw}}=1,\ldots,1024$) and dense regime ($N_{\mathrm{cw}}=2048,\ldots,20480$) are visible in a single panel. \textbf{Right:} DGX Spark throughput versus decoder iterations, aggregated over the dense regime $N_{\mathrm{cw}}\in\{2048,4096,\ldots,20480\}$; text labels indicate GB10/Grace throughput speedup. The left panel exposes three operating regions (launch-limited small batches, GPU ramp-up, and dense steady-state operation), while the right panel preserves the dense-regime edge-RSU view underlying the paper's main claim.}}
  \label{fig:throughput_scaling}
\end{figure*}

Figure~\ref{fig:throughput_scaling} provides throughput analysis by separating \emph{full-range hardware scaling} from the \emph{dense-regime DGX Spark view}. Because the expanded sweep spans more than four orders of magnitude in batch size, the left panel uses a $\log_2$ horizontal axis; equal spacing therefore corresponds to doubling $N_{\mathrm{cw}}$, which makes the small-batch transition region visible without obscuring dense-regime behavior. The left panel aggregates over decoder iterations for each $N_{\mathrm{cw}}$ and shows three distinct operating regions. First, at very small batch sizes, CPUs are competitive with or faster than GPUs because launch and setup overhead dominate useful parallel work. On DGX Spark, the Grace CPU remains ahead through $N_{\mathrm{cw}}=8$, and the GB10 GPU overtakes by $N_{\mathrm{cw}}=16$. On the COTS platform, the i9-14900K CPU and RTX~4090 GPU are near parity at $N_{\mathrm{cw}}=4$, and the GPU overtakes by $N_{\mathrm{cw}}=8$. This clarifies that GPU acceleration is not universal at tiny batches; the “Six Times to Spare” result applies to the dense regime, not a statement about all $N_{\mathrm{cw}}$ values.

Second, both platforms exhibit a clear GPU ramp-up region as batch parallelism increases. On DGX Spark, the GPU reaches roughly $2.0\times$ throughput speedup by $N_{\mathrm{cw}}=32$ and $3.72\times$ by $N_{\mathrm{cw}}=128$. On COTS, the RTX~4090 reaches approximately $2.44\times$ by $N_{\mathrm{cw}}=64$, $3.48\times$ by $N_{\mathrm{cw}}=128$, and $5.46\times$ by $N_{\mathrm{cw}}=256$. Because the left panel is plotted on a $\log_2$ scale, this ramp-up should be read as a progression across successive doublings of batch size rather than as a linear-in-$N_{\mathrm{cw}}$ slope.

Third, both platforms enter a steady-state dense regime. On DGX Spark, the GB10/Grace speedup settles to approximately $5.82\times$ throughout $N_{\mathrm{cw}}=2048\ldots 20480$. On the COTS platform, the RTX~4090 already reaches about $13.5\times$ by $N_{\mathrm{cw}}=1024$ and then remains in the approximate $15.0\times$--$15.8\times$ range across $N_{\mathrm{cw}}=2048\ldots 20480$. The logarithmic axis makes clear that, once the small-batch transition is passed, the dominant visual effect is not continued rapid scaling with batch size but convergence toward a platform-dependent steady-state throughput region. The full-range sweep explains both \emph{where} GPU acceleration emerges and \emph{why} the DGX Spark dense regime is the correct operating region for the paper's edge-provisioning claim.

The right panel of Fig.~\ref{fig:throughput_scaling} focuses on the DGX Spark's optimal dense-regime operation for edge-RSU deployment. Throughput falls as the belief-propagation iteration count increases, as expected for an iterative message-passing decoder whose work grows approximately linearly with $I$. Even so, the GB10 curve remains consistently above the Grace curve across the full range. At $I=4$, mean dense-regime throughput is approximately $19.5\,\mathrm{Mbit/s}$ on GB10 versus $3.35\,\mathrm{Mbit/s}$ on Grace; by $I=22$, it decreases to roughly $3.7\,\mathrm{Mbit/s}$ versus $0.65\,\mathrm{Mbit/s}$. The separation is remarkably stable: dense-regime mean speedup remains in the narrow $\sim 5.7\times$--$5.8\times$ range across all tested iteration budgets. DGX Spark's platform precision is practically useful for edge-RSU reasoning: it does not merely accelerate one favorable point, but preserves a similar factor-of-six decoder advantage across a realistic span of decoding effort.

A useful systems-level contrast emerges when the full-range scaling is interpreted through amortized service time rather than throughput alone. The $\log_2$ presentation of $N_{\mathrm{cw}}$ is particularly helpful here because it emphasizes the transition from launch-limited execution to near-constant amortized service time across successive doublings of batch size. On DGX Spark, the GB10 GPU reaches an approximately steady per-codeword service time by about $N_{\mathrm{cw}}\approx 256$, after which additional batch growth primarily sustains rather than materially improves the amortized rate. The RTX~4090, by contrast, continues improving deeper into the sweep before settling, reflecting the larger discrete accelerator's ability to absorb additional parallel work. That higher absolute performance, however, does not make the workstation the better \emph{edge} result: the COTS platform achieves its advantage with a discrete, PCIe-attached, higher-power GPU and a non-integrated memory hierarchy, whereas DGX Spark delivers its dense-regime headroom on a coherent CPU+GPU SoC intended for compact edge deployment.

The localized, temporary RTX~4090 peak in Fig.~\ref{fig:throughput_scaling_codewords} reveals deeper architectural truth. To illustrate this, we performed a focused COTS spike ablation over $N_{\mathrm{cw}}\in\{512,1024,2048\}$ and $I\in\{4,10,20\}$ (19 profiled runs), summarized in Table~\ref{tab:cots_spike_ablation}. The local maximum near $N_{\mathrm{cw}}\approx 1024$ is real and replicated: at $I=4$, repeated 1024-codeword peak runs reached $158.23$--$160.72$\,Mbit/s on the RTX~4090, versus $116.61$--$117.79$\,Mbit/s at $N_{\mathrm{cw}}=512$ and $126.94$--$130.42$\,Mbit/s at $N_{\mathrm{cw}}=2048$. However, the post-peak decline cannot be attributed to hard PCIe saturation: across all profiled runs the 4090 remained at Gen4 $\times$16 current=max, while observed PCIe Rx/Tx stayed only in the tens to low hundreds of MB/s, far below a stressed link. Instead, TensorFlow device-placement logs and process telemetry show a strongly hybrid CPU+GPU decoder path whose hottest dense runs consume $\sim 27$--$29$ active-core equivalents on the host during GPU runs to orchestrate system RAM$\leftrightarrow$GPU VRAM copies, indicating that the 1024-codeword point is a medium-batch efficiency sweet spot in which GPU occupancy is high while decoder workspace and runtime orchestration remain relatively favorable. By 2048 and beyond, the larger internal working set moves the implementation into a lower-efficiency, runtime-heavier dense regime. This statistical phenomena is not to be confused with the $10-12$ active core offload reported for DGX Spark, whose less pronounced throughput saturation can be attributed to coherent NVLink-C2C CPU$\leftrightarrow$GPU memory architecture. This removes the discrete host/device boundary and eliminates the need for CPU ``middle management'' orchestration, producing the smoother GB10 ramp and earlier steady-state behavior seen in Fig.~\ref{fig:throughput_scaling_codewords}.

\begin{table}[t]
  \centering
  \scriptsize
  \setlength{\tabcolsep}{4pt}
  \caption{{\small Focused COTS spike ablation around the RTX~4090 local maximum in Fig.~\ref{fig:throughput_scaling_codewords}. Throughput values show the observed range across repeated profiled runs at the selected $(N_{\mathrm{cw}},I)$ points.}}
  \label{tab:cots_spike_ablation}
  \begin{tabular}{c c c c c}
    \toprule
    $N_{\mathrm{cw}}$ & $I$ & i9 throughput & RTX~4090 throughput & Speedup \\
     &  & [Mbit/s] & [Mbit/s] &  \\
    \midrule
    512  & 4  & 16.07--16.10 & 116.61--117.79 & 7.26--7.32$\times$ \\
    1024 & 4  & 12.68--12.70$^{\dagger}$ & 158.23--160.72$^{\dagger}$ & 12.46--12.68$\times$ \\
    2048 & 4  & 8.30--8.67 & 126.94--130.42 & 15.05--15.30$\times$ \\
    512  & 20 & 3.45--6.74 & 31.68--61.10 & 9.06--9.95$\times$ \\
    1024 & 20 & 2.60--2.68 & 36.28--37.71 & 13.96--14.05$\times$ \\
    2048 & 20 & 1.72--3.51 & 27.03--54.85 & 15.48--16.29$\times$ \\
    \bottomrule
  \end{tabular}
  \vspace{2pt}
  \begin{flushleft}
  \footnotesize Across 19 focused COTS runs, the RTX~4090 remained at Gen4 $\times$16 current=max, observed PCIe Rx/Tx stayed in the tens to low hundreds of MB/s, and peak Python CPU load reached $\sim 27$--$29$ active-core equivalents in the hottest dense runs. $^{\dagger}$A third profiled run at $(1024,4)$ recorded 5.20/70.42\,Mbit/s (i9/4090) with the same link state, consistent with host-runtime interference rather than PCIe throttling.
  \end{flushleft}
  \vspace{-15pt}
\end{table}

To connect these throughput results to URLLC provisioning intuition, Table~\ref{tab:per_cb_latency} reports the mean \emph{amortized} per-codeword service time $t_{\mathrm{cb}} \triangleq \frac{t_{\mathrm{dec}}}{N_{\mathrm{cw}}}$ for representative iteration counts in the dense regime, normalized by a nominal NR slot budget of $0.5$\,ms. On DGX Spark, GB10 requires only $0.026$\,ms, $0.064$\,ms, and $0.126$\,ms per codeword at $I\in{4,10,20}$, corresponding to $5\%$, $13\%$, and $25\%$ of the slot budget. Grace, in contrast, consumes $31\%$, $75\%$, and $145\%$ of that same budget. This is the core edge-platform result: in the dense operating region, GPU offload transforms LDPC decoding from a potentially slot-dominating CPU kernel into a substantially smaller component of the edge compute budget. The COTS platform is faster in absolute terms, especially on RTX~4090, but we interpret that as an upper-bound comparison rather than the deployment recommendation.

\begin{table}[t]
  \centering
  \scriptsize
  \setlength{\tabcolsep}{3.0pt}
  \caption{{\small Mean amortized per-codeword LDPC decode time $t_{cb}=t_{dec}/N_{cw}$ versus a 0.5\,ms NR slot in the dense regime, averaged over $N_{cw}\in\{2048,4096,\dots,20480\}$ (10 batch sizes) and 10 repetitions per point (100 summarized samples/row). DGX Spark uses a Grace ARM CPU and GB10 GPU; COTS PC uses an i9-14900K CPU and RTX 4090 GPU.}}
  \label{tab:per_cb_latency}
  \begin{tabular}{c cc cc cc cc}
    \toprule
    & \multicolumn{4}{c}{DGX Spark} & \multicolumn{4}{c}{COTS PC} \\
    \cmidrule(lr){2-5}\cmidrule(lr){6-9}
    Iter &
    \makecell{Grace \\ (ms)} &
    \makecell{Grace / \\ 0.5\,ms} &
    \makecell{GB10 \\ (ms)} &
    \makecell{GB10 / \\ 0.5\,ms} &
    \makecell{i9-14900K \\ (ms)} &
    \makecell{i9 / \\ 0.5\,ms} &
    \makecell{RTX 4090 \\ (ms)} &
    \makecell{4090 / \\ 0.5\,ms} \\
    \midrule
    4  & 0.153 & 0.31 & 0.026 & 0.05 & 0.087 & 0.17 & 0.006 & 0.01 \\
    10 & 0.373 & 0.75 & 0.064 & 0.13 & 0.213 & 0.43 & 0.015 & 0.03 \\
    20 & 0.725 & 1.45 & 0.126 & 0.25 & 0.423 & 0.85 & 0.029 & 0.06 \\
    \bottomrule
  \end{tabular}
\end{table}

\begin{figure}[t]
  \centering
  \includegraphics[width=\columnwidth]{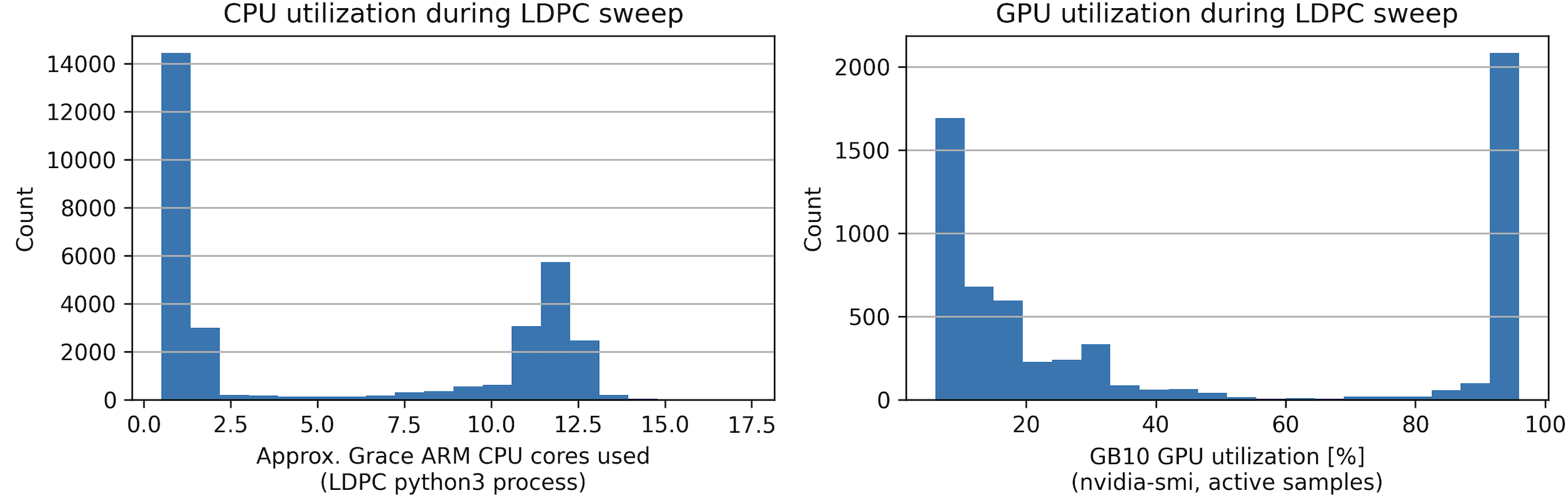}
  \caption{{\small Resource usage during LDPC5G decoding on DGX Spark. Left: histogram of approximate Grace cores used by the LDPC process. Right: histogram of GB10 GPU utilization for active samples (utilization $>5\%$), as reported by \texttt{nvidia-smi}.}}
  \label{fig:resource_usage}
\end{figure}

Compute headroom is only operationally meaningful if it coincides with visible resource rebalancing on the target edge platform. Figure~\ref{fig:resource_usage} summarizes DGX Spark telemetry captured during the sweep. The left panel shows the approximate number of active Grace CPU cores consumed by the LDPC process, while the right panel shows GB10 utilization during active decode periods. CPU-based execution commonly occupies on the order of $\sim 10$--$12$ active-core equivalents during busy intervals, whereas GPU-based execution drives the accelerator to high occupancy, frequently above $90\%$ utilization in active samples. The practical takeaway is that the throughput and service-time improvements are accompanied by a real shift in where the work is performed: LDPC decode pressure moves away from the general-purpose CPU budget and onto the accelerator, which is exactly the rebalancing desired on an edge node concurrently hosting scheduling, control, and higher-layer vehicular workloads.

Power telemetry further distinguishes the two platforms. On DGX Spark, high-occupancy GB10 decode windows cluster near $\sim 30$\,W from an $\sim 11$--$12$\,W low-utilization baseline, consistent with a modest incremental accelerator cost. On the COTS system, after benchmark initialization, the RTX~4090 sits near a $\sim 62$\,W low-utilization baseline with the benchmark resident and rises to $\sim 145$\,W across active samples ($>5\%$ GPU utilization) and $\sim 220$\,W in $>80\%$-utilization windows, i.e., $80$--$160$\,W above baseline depending on occupancy. This reinforces why the COTS platform is best interpreted as an upper-bound comparison rather than an edge deployment target.

\section{Implications for URLLC Vehicular Edge Base Stations}
\label{sec:discussion}
DGX Spark remains the deployment-relevant result in this study because it is the compact heterogeneous edge node: a coherent CPU+GPU platform intended for colocated RAN and edge workloads under practical footprint and power constraints. By contrast, the COTS i9-14900K/RTX~4090 upper-bound illustrates what a large discrete GPU can deliver when edge constraints are relaxed, including a much larger load-dependent accelerator power increase. We provide upper-bound context and edge-aware deployment insight

The central implication of Section~\ref{sec:results} is that GPU offload can materially change the \emph{compute shape} of an edge-RSU baseband pipeline on the target platform. In the dense operating regime, GB10 reduces the amortized per-codeword service time $t_{\mathrm{cb}} = t_{\mathrm{dec}}/N_{\mathrm{cw}}$ to a small fraction of a representative $0.5$\,ms slot budget: approximately $5\%$, $13\%$, and $25\%$ at $I\in\{4,10,20\}$, respectively (Table~\ref{tab:per_cb_latency}). The corresponding Grace-only values are substantially higher at $31\%$, $75\%$, and $145\%$. For an edge base station, this changes LDPC decoding from a potential ``slot dominator'' on the CPU into a bounded accelerator-resident component, increasing the likelihood that the full PHY budget---including demapping, HARQ handling, scheduler interaction, and adjacent signal-processing stages---can be met within URLLC-oriented timing envelopes~\cite{popovski2018urllc_building_blocks,qualcomm2017_urllc_systems}.

From an RSU/gNB dimensioning perspective, the dense-regime DGX Spark throughput advantage in Fig.~\ref{fig:throughput_scaling} translates into \emph{higher concurrency headroom} for code blocks. The batch dimension $N_{\mathrm{cw}}$ is an abstraction of simultaneous decode demand arising from multiple UEs, multiple HARQ processes, or transiently heavy scheduling intervals. If the decoder consumes a smaller fraction of the slot budget and maintains a stable factor-of-six throughput advantage across iteration counts, then the node can absorb larger bursts of decode demand before timing margins are exhausted. This is particularly relevant in vehicular settings, where intersection dynamics and bursty offered load can create short periods of unusually high concurrent PHY demand.

The system-level value of this headroom is magnified when the same RSU/edge node must host workloads \emph{beyond} the radio stack. Figure~\ref{fig:resource_usage} shows that accelerated decoding shifts substantial active work away from the general-purpose Grace CPU budget and onto GB10, relieving multi-core CPU pressure during LDPC service. That relief is not merely an implementation detail: reclaimed CPU cycles can reallocate to latency-sensitive control-plane functions, scheduler logic, protocol handling, and colocated vehicular edge applications such as cooperative perception or fusion pipelines~\cite{zhang2021emp}. GPU offload is valuable not only because it makes decoding faster, but because it rebalances scarce edge-node resources in architecturally useful directions for RSU deployments~\cite{liu2021vec_survey}.

At the same time, the batching interpretation must be kept precise. The reported quantity $t_{\mathrm{cb}}=t_{\mathrm{dec}}/N_{\mathrm{cw}}$ is an \emph{amortized service time under batch-parallel execution}, not a direct measurement of single-codeword arrival-at-idle latency. It is intentionally evaluated at large $N_{\mathrm{cw}}$ in the dense regime to expose steady-state efficiency and decoder headroom under heavy parallel load. It is a useful provisioning metric, but not a complete latency metric. Real deployments may operate at smaller instantaneous batch sizes, especially outside congested intervals, and in that region constant overheads such as framework dispatch, synchronization, queueing, and launch costs become more visible. The baseline regime in Fig.~\ref{fig:throughput_scaling} makes this explicit: at very small batches, CPUs remain competitive until a GPU has enough parallel work to amortize fixed costs. Future work should measure p95/p99/p99.99 decode latency and end-to-end slot-budget composition under contention, including interactions with the full PHY, HARQ feedback timelines, scheduler decisions, and concurrent edge workloads.


Within those bounds, the deployment takeaway is straightforward. For compact heterogeneous edge nodes such as DGX Spark, accelerator offload can reshape LDPC decoding from a CPU-bound bottleneck into a relatively small and predictable consumer of the slot budget, while simultaneously freeing CPU cores for the rest of the base station and colocated vehicular intelligence. The GPU is not only faster, it creates edge-RSU URLLC advantage via a surplus of compute headroom, and empowers adjacent edge functions on the same node.

\section{Conclusion: Six Times to Spare for Edge-RSU LDPC}
\label{sec:conclusion}

This paper presents a reproducible, telemetry-backed LDPC5G decode microbenchmark derived from the Sionna baseline to quantify decoder headroom on compact heterogeneous edge hardware. On DGX Spark, in the dense regime $N_{\mathrm{cw}}\in\{2048,4096,\ldots,20480\}$, GB10 sustains an approximately $5.8\times$ throughput advantage over Grace across the tested iteration range, supporting the \emph{``Six Times to Spare''} edge claim: against a representative $0.5$\,ms slot budget, the CPU can approach or exceed the slot at high iteration counts, while the GPU remains a small fraction of it. The full ablation over $N_{\mathrm{cw}}=1\ldots 20480$ also defines the boundary of this result: CPUs remain competitive at very small batches, GPUs gain advantage as parallelism rises, and only then enter the dense steady-state regime supporting the edge claim. Although the COTS i9-14900K/RTX~4090 platform reaches a higher absolute ceiling under relaxed constraints, it is best treated as an upper bound rather than an edge-suitable point. For compact URLLC vehicular edge nodes, DGX Spark shifts LDPC decoding from a CPU bottleneck to a bounded accelerator-resident workload, freeing CPU headroom for the rest of the radio stack and colocated edge intelligence.

The full-range scaling study also reveals an architectural lesson beyond raw throughput: on discrete PCIe GPUs, scaling artifacts can arise not from exhausted GPU compute, but from CPU orchestration overhead and system RAM$\leftrightarrow$VRAM transfers across the host--device boundary. In our COTS ablation, this appears as a medium-batch sweet spot followed by heavier runtime behavior at larger working sets. By contrast, coherent edge memory architectures such as DGX Spark's NVLink-C2C fabric reduce this host-side penalty by removing the discrete RAM/VRAM split and enabling a smoother path to steady-state operation. For edge-RSU design, this suggests coherent memory can matter as much as peak accelerator throughput when predictable PHY headroom is the real objective.

\bibliographystyle{IEEEtran}
\bibliography{References}

\end{document}